\documentstyle[epsfig,aps,preprint,tighten]{revtex}

\begin{document}

\large\normalsize

\title
{Cuts in the invariant mass of resonances in many body decays of
mesons}
\author
{L.N. Epele, D. G\'omez Dumm, A. Szynkman}

\address
{IFLP, Depto.\ de F\'{\i}sica, Universidad Nacional de La Plata,
C.C.\ 67, 1900 La Plata, Argentina}

\author
{R. M\'endez--Galain}

\address
{Instituto de F\'{\i}sica, Facultad de Ingenier\'{\i}a, Univ. de la
Rep\'ublica, C.C.\ 30, CP 11000 Montevideo, Uruguay}

\maketitle

\begin{abstract}
Resonance-mediated many body decays of heavy mesons are analyzed. We
focus on some particular processes, in which the available phase space
for the decay of the intermediate resonance is very narrow. It is shown
that the mass selection criteria used in several experimental studies of
$D$ and $B$ meson decays could lead to a significant underestimation of
branching ratios.
\end{abstract}



\vspace{1cm}

Three or more body decays of heavy mesons are usually dominated by
intermediate resonances. The latter are short lived states that
cannot be directly observed: only daughter particles produced
through their decays reach the detectors\cite{PDG}. The detected
final state results from the interference of all possible
intermediate channels. Thus, one has to disentangle the quantum
interference to understand the underlying physics. A powerful
technique to split up the various resonant channels is the
so-called Dalitz plot fit\cite{DP}.

In general, if a given resonance proceeds within a kinematic region
where no other resonances occur, interference effects are assumed to be
negligible. In this case, the branching ratio of the heavy meson decay
to that particular resonance can be measured in a simple way: one just
counts the number of events in which the invariant mass of the decay
products lies within a small window around the resonance central mass.
The size of this window is chosen according to the width of the
resonance. This is a natural and thus widely used measurement
technique\cite{ventana}.

In this Letter we discuss the validity of this method. In fact, the
resonance is a virtual state, and its squared four-momentum can reach in
principle any value within the allowed phase space. This is certainly
not a new statement. The interesting fact is that, as we will see, for
some particular decays one could find a very large number of (naively,
unexpected) events such that, even if they proceed through a particular
resonance, the invariant mass of the detected particles is indeed very
far from the resonance mass shell (with ``far'', we mean in comparison
with the resonance width). These events would be missed if the counting
method simply considers a narrow window around the resonance central
mass. As shown below, this is the case for some processes in which the
resonance decay rate is kinematically suppressed. We are aware that the
physics involved in this discussion is well known. However, we believe
that the magnitude of the effect has not been well appreciated so far,
and the described counting method is not always safely applicable. Here
we present simulations of actual decays and discuss the significance of
the effect as well as the expected experimental difficulties.

Let us first describe those aspects of many body decays relevant to our
discussion. Consider a heavy meson $P$ decaying into a final state given
by three detected particles, $d_1, d_2$ and $d_3$, and assume that the
decay proceeds through intermediate resonances $R_1, R_2$,
etc.\footnote{The decay also proceeds through a non-resonant channel,
which is not relevant to our discussion.}. The total amplitude of the
decay is the sum of the amplitudes of all partial channels, each one
mediated by a resonance $R_i$:
\begin{equation}
{\cal A}_{tot} = \Sigma_i {\cal A}_{R_i}\;.
\label{ampltot}
\end{equation}

For simplicity, let us assume that the decay is dominated by a single
resonance $R$, in such a way that ${\cal A}\simeq {\cal A}_R = {\cal A}(P
\to R d_3; R \to d_1 d_2)$. It is natural to describe the process by
considering three independent stages: resonance production $P \to R d_3$,
resonance propagation, and resonance decay $R \to d_1 d_2$. The amplitude
factorizes then as
\begin{eqnarray}
{\cal A}(P \to R d_3 \to d_1 d_2 d_3) & =
\nonumber \\
{\cal A}(P \to R d_3) \times
BW_{R,12} \times {\cal A}(R \to d_1 d_2) \;. &
\label{fact}
\end{eqnarray}

The amplitudes ${\cal A}(P \to R d_3)$ and ${\cal A}(R \to d_1 d_2)$ have to
take into account the conservation of angular momentum in the decays, as
well as the energy dependence, usually parameterized through form factors
(we will come back to these two ingredients later). As usual, we describe
the resonance propagation by means of a relativistic Breit-Wigner
function\cite{BW}
\begin{equation}
BW_{R,12} (m_{12}^2) = \frac{1}{m_0^2 - m_{12}^2 - i m_0 \Gamma}\; ,
\label{BW}
\end{equation}
where $m_0$ is the resonance mass, $\Gamma$ is the resonance width, and
$m_{12}^2$ is the invariant mass of the outgoing particles $d_1$ and $d_2$,
$m_{12}^2 =(p_1 + p_2)^2$.

The function $|BW(m_{12}^2)|$ is peaked around the resonance mass, and
decreases with a rate given by $\Gamma$. This behavior reflects the fact
that $R$ is a virtual particle that ---according to quantum mechanics--- can
have {\em any} invariant mass $m_{12}^2$, the relative probability of each
``mass'' being weighted by the factor $|BW(m_{12}^2)|^2$. It is then natural
to measure $BR(P \to R d_3)$ by simply counting the number of detected
particles $d_1$ and $d_2$ for which the value of $\sqrt{m_{12}^2}$ lies
within a region of the order of $(m_0-\Gamma,m_0+\Gamma)$. This technique
has in fact been used for many years\cite{ventana}. However, as we will show
in the following, the usage of this approach is not always safe.

The decay of a resonance is usually driven by the strong interaction. Thus,
resonances have relatively large widths ---some dozen or even some hundreds
of MeVs~\cite{PDG}. However, if a particular decay is somehow suppressed,
the resonance may have a longer life, and its width can be as small as some
MeVs or less. This happens in particular when the resonance central mass
$m_0$ is very close to the threshold of its decay to $d_1 d_2$. In this
case, the phase space available for the decay turns out to be very narrow,
and the (virtual) resonance could be allowed to decay through other channels
which were in principle expected to be strongly suppressed in comparison
with the ``natural'' strong channel $R\to d_1 d_2$.

This is the case, for instance, of the $\phi(1020)$ vector meson. For this
resonance, the natural decay channel is $K \bar K$, in the same way as the
natural decay for $\rho$ is two pions. Nevertheless, due to the small phase
space available ---32 MeV and 24 MeV for the charged and neutral kaons,
respectively--- the corresponding $\phi$ branching ratios are ``only'' 49\%
and 34\% for $K^+K^-$ and $K \bar K$, respectively. Electromagnetic decays,
that have branching fractions as small as $10^{-4}$ in the $\rho$ decay
pattern, are of the order of 1\% in the $\phi$ case. Accordingly, the $\phi$
width is about 35 times smaller than the $\rho$ one.

Other examples are low mass $D^\ast$ vector mesons. Their natural decay
channel is $D \pi$, in the same way as the natural decay for $K^\ast$ is $K
\pi$. But here, the phase space is as small as 7 MeV for the
${D^\ast}^0(2007) \to D^0 \pi^0$, 6 MeV for both ${D^\ast}^+(2010)\to D^0
\pi^+$ and ${D^\ast}^+(2010)\to D^+ \pi^0$, and 8 MeV for ${D_s^\ast}^+\to
D_s^+ \pi^0$. As a consequence, measured branching ratios of these decays
---which otherwise should reach almost 100\%--- are as ``small'' as 62\%,
68\%, 31\% and 6\%, respectively\footnote{$D_s^{+*} \to D_s^+ \pi^0$ is
an isospin violating decay (see Ref.\ \cite{dsdspi}). It thus has
another, phase space independent suppression.}. In the
${D^\ast}^0(2007)$ decay pattern, the electromagnetic decay is as large
as 38\%, i.e., of the same order of magnitude of the strong one (in
contrast, in the $K^\ast$ case, electromagnetic decays are of the order
of $10^{-3}$). Accordingly, resonance widths are quite small: the width
of ${D^\ast}^+(2010)$ has recently been reported to be as small as 0.1
MeV\cite{cleo}, whereas for ${D^\ast}^0(2007)$ and ${D_s^\ast}^+$ only
upper limits are known, presently of the order of 2 MeV.

Let us face the study of heavy meson decays mediated by these particular
spin one resonances, focusing on cases in which the detected final state
includes their natural, and yet highly suppressed, strong channels. We
describe here a usual situation\cite{PDG}, where both the initial heavy
meson and the particles in the final state are scalars, $P =D, B$, and $d_i
= \pi, K, D$. In this case, Eq. (\ref{fact}) can be conveniently written as
\begin{equation}
{\cal A}(P \to R d_3 \to d_1 d_2 d_3)   =
F_{P,Rd_3}\; F_{R,d_1d_2} \; (-2 {\vec{p}_1}\cdot{\vec{p}_3}) \; BW_{R,12}\;,
\label{vector}
\end{equation}
where $F_{P,Rd_3}$ and $F_{R,d_1d_2}$ are form factors, and the
three-momenta ${\vec{p}_1}$ and ${\vec{p}_3}$ are evaluated in the
resonance rest frame. The explicit momentum dependence in
(\ref{vector}) follows from Eq.\ (\ref{fact}), just assuming
Lorentz invariance and summing over all possible polarizations of
the intermediate vector meson resonance. The differential decay
width of this reaction can be written as
\begin{equation}
d\Gamma  = \frac{1}{(2\pi)^3}\frac{1}{32M^3}|{\cal A}|^2
dm^2_{12}\, dm^2_{13} \;,
\label{width}
\end{equation}
where $m^2_{ij}=(p_i+p_j)^2$ and $M$ is the mass of the decaying meson $P$.

For the decays considered here, there is a strong suppression at the
resonant peak, i.e.\ when $m_{12}^2\simeq m_0^2$. This suppression is due to
purely kinematic effects. To see this, let us take ${\cal A}\simeq$ constant
for a given value of $m_{12}^2$ ---which means to neglect the dynamics of
the decay--- and integrate over the variable $m^2_{13}$ within the kinematic
limits of the three body phase space. It is easy to see that
\begin{equation}
\frac{d\Gamma}{dm^2_{12}}  \propto |{\cal A}|^2\,|{\vec{p}_1}| \;,
\label{diffwidth}
\end{equation}
where ${\vec{p}_1}$ is the three-momentum of $d_1$ in the resonance rest
frame. Since we are assuming that the mass of the resonance is just above
the threshold, $m_0\simeq m_1+m_2$, at the resonance peak both particles
$d_1$ and $d_2$ will be produced almost at rest. Thus Eq.\ (\ref{diffwidth})
implies a suppression in the partial width.

The effect is even stronger in the particular case of a vector
resonance. In that case, as stated in Eq.\ (\ref{vector}), the
amplitude is proportional to ${\vec{p}_1} \cdot {\vec{p}_3}$.
Assuming that the form factors are slow-varying functions of the
phase space variables, the partial width is finally expected to be
suppressed by a factor $(|\vec p_1|/\Lambda)^3$, where $\Lambda$
is some natural scale of the process, typically of order $M$.

Eqs.\ (\ref{vector}) and (\ref{diffwidth}), which are certainly very
well known, show up the main point of this Letter. The decay width
$\Gamma(P\to d_1 d_2 d_3)$ is driven by two {\it competitive} effects.
On the one hand, the BW propagator strongly enhances the decay amplitude
in the vicinity of the resonance mass. On the other hand, kinematic
effects suppress the differential width at the resonance peak, hence the
decay rate for other kinematic regions is comparatively enhanced. The
usual suppression of the differential decay width for $P \to R d_3 \to
d_1 d_2 d_3$ outside the window allowed by the BW function is not
obvious in this case.

To clarify our point, let us emphasize that we are not claiming that there
is an enhancement of the resonant {\it production} probability $P \to R d_3$
outside the BW peak. On the contrary, the point is that due to the
suppression of the resonance {\it decay rate} $R \to d_1 d_2$ at the
resonance central mass, the {\it combined} production + decay probability
within and outside the peak could be of comparable orders. In other words,
the total width (i.e., integrated over the whole phase space) is
indeed small; the important point is that, even if the decay proceeds
through a BW-described resonance, the relative weight for the decay
rate near or far from the resonance mass shell is not just driven by
the BW function.

In order to show that this effect is not a simple academic thought, we
present in the following a simulation of an actual process.
We first present the pure theoretical estimate; afterwards, we will consider
the experimental difficulties.
Let us consider the decay $B^+ \to \bar D^{\ast 0} D^+_s ;
\bar D^{\ast 0}\to \bar D^0\pi^0$. Using Eq.\ (\ref{vector}), it is
possible to get an estimate for the differential decay rate $d\Gamma$ as
a function of $m_{12}^2$. Since the form factors are usually smooth
functions, we will assume as a first guess that they are constant
(experimental analyses show that form factor shapes have no significant
effect on the total systematic error of a given Dalitz plot
fit\cite{exp}). In this way we can calculate the ratio
\begin{equation}
r  = \frac{\displaystyle
\int_{m^2_{12}=(m_0-n\,\Gamma)^2}^{m^2_{12}=(m_0+n\,\Gamma)^2}
|{\cal A}|^2 d\Phi }{\displaystyle \int |{\cal A}|^2 d\Phi }\;,
\label{r}
\end{equation}
where $d\Phi$ is an element of the three body phase space, $m_0$ is the
$\bar D^{\ast 0}$ resonance mass ($m_0 = 2007$ MeV), $\Gamma$ is the
$\bar D^{\ast 0}$ width, and $n$ is a real number. $\Gamma$ is presently
unknown, its upper limit being 2.1 MeV with a 90\% confidence
level\cite{PDG}. The integral in the denominator is calculated over the
whole phase space, while that in the numerator is limited to a window in
$m^2_{12}$. Thus, $r$ is a measure of the relative number of events that
are expected to fall within the resonance peak.

We quote in Table I the values of $r$ for some input values of $n$ and
$\Gamma$. Our results show that the effect we are describing can
be very strong if the resonance width is as large as 2 MeV, and it
remains quite significant even if $\Gamma$ is of the order of 0.1 MeV.
For comparison, we include in the last column the values of $r$
corresponding to a fictitious resonance having the same quantum numbers
as $\bar D^{\ast 0}$ but a higher mass, $m_0 = 2.6$~GeV. This particle
would not suffer the kinematic suppression in its decay to $\bar
D^0\pi^0$, consequently a small value of $n$ is enough to get $r$ above
90\%. We have found that in this case the results for $r$ are
independent of the specific value of $\Gamma$.

Figures 1a and 1b show the kinematic distribution of the events for the
three body decay $B^+ \to \bar D^0 \pi^0 D^+_s$, assuming that the
process is dominated by the $\bar D^{\ast 0}(2007)$ resonance
channel\footnote{In fact, the decay can also proceed through an
intermediate resonance ${D_s^\ast}^+$, as well as other higher resonant
states. The inclusion of these contributions in our analysis does not
affect significantly the results.} and considering a $\bar D^{\ast 0}$
width $\Gamma = 1$ MeV. The plots correspond to a Monte Carlo simulation
of 10000 events, performed using Eqs.\ (\ref{vector}) and (\ref{width}).
Figure 1a is the Dalitz plot of the decay as a function of the invariant
masses $m^2_{\bar D^0\pi^0}$ and $m^2_{D^+_s\pi^0}$, while in Figure 1b
we represent an histogram of the number of events as a function of
$m^2_{\bar D^0\pi^0}$.

It is seen that the Dalitz plot shape in Fig.\ 1a is quite different
from that naively expected for a decay mediated by a $\Gamma = 1$ MeV
vector resonance. This becomes evident by looking at Figure 1c, where we
show a Monte Carlo simulation of the same process, now shifting the
$\bar D^{\ast 0}$ mass to the fictitious value of 2.6~GeV considered in
Table I. The striking difference between the event distribution in both
plots arises from the kinematic suppression discussed above. However,
the situation displayed in Fig.\ 1a can be misleading if one just looks
at the events for which $m^2_{\bar D^0\pi^0}$ is near the $\bar D^{\ast
0}$ mass. In fact, despite the spreading of events along the whole phase
space, the event density is still much larger in the peak region than
anywhere else. Therefore, Figure 1b could be mistaken for a $\Gamma = 1$
MeV Breit-Wigner function, with some background in the right
sideband\cite{prl}. It would be then natural to apply the usual method,
that means to consider that the events within a window of a few $\Gamma$
around the peak amount almost the total number of decay events. But,
according to our simulation, this would be wrong by a factor as large as
3 to 4 (see Table I). Indeed, even if the amplitude is peaked around the
resonance mass, the phase space area outside the peak is comparatively
so large that the number of events falling in this region becomes very
important.

Let us now discuss the experimental difficulties that could appear when
data from real events are analyzed. First of all, resolution in these
experiments is usually larger than the width of the intermediate
resonances involved\cite{ventana}. As a consequence, it would be
impossible to access to a direct measurement of the ratio $r$, at least
for small values of $n$. Accordingly, the experimental histogram of
Fig.\ 1b would be wider, arising from the convolution of a narrow
Breit-Wigner function (physical) and a Gaussian function with a larger
width (resolution). Nevertheless, the bulk of our reasoning remains
valid: a large amount of physical events may fall outside the peak. A
second comment refers to the difficulty of developing a more careful
analysis of data, in order to include the contribution of these events.
Indeed, according to Table I, the amount of events could be quite
important, but at the same time they appear to be spread in a very large
region of the phase space. Then, only a relatively low number of events
per bin would be found outside the peak, and they could be hardly
disentangled from the background, no matter which is the function used
to perform the corresponding fit.

It is important to stress that our theoretical estimates rely on two
basic assumptions. First, we have taken the form factors to be
approximately constant along the phase space. Second, we have assumed
that the Breit-Wigner shape remains valid far beyond a small region
around the peak. Whereas the first hypothesis is quite natural and does
not have a significant effect on our results, the second assumption can
be hardly supported from the theoretical point of view. Our simulations
are in this sense strongly model dependent, and this has to be kept in
mind when looking at the numerical values presented in Table I.

In any case, our simulations can be taken as a severe warning,
indicating that many reported results can be spoiled by the effect
described in this Letter. For some particular processes, the magnitude
of the corrections could be quite significant, and this should be taken
into account ---at least--- in the corresponding systematic errors. This
includes the analysis of the decays $B \to \bar D^{\ast 0} d_3$, $B\to
{D^\ast}^+ d_3$ and $B \to {D_s^\ast}^+ d_3$, with $d_3 = \pi^0, \eta,
K, D_s^+, D^0$, etc. These channels have been measured\cite{lista} using
$\bar D^0 \pi^0 d_3$ ---respectively $D^+ \pi^0 d_3$ and $D_s^+ \pi^0
d_3$--- as final states, and imposing a cut in the invariant mass
$m_{D^0 \pi^0}^2$
---respectively $m_{D^+ \pi^0}^2$ and $m_{D^+_s\pi^0}^2$. According to
the simple analysis presented here, in all these cases the effect would
be of the order predicted in Table I. In particular, in the case of the
${D^\ast}^+$ resonance, whose width has recently been reported to be 0.1
MeV, the effect is expected to be of the order of 30\%.

Finally, let us mention that $\phi$ meson production and decay
measurements could also be affected by the effect described above. Since
the $\phi$ resonance can be produced in $D$ meson decays, a significant
amount of data is presently available, and many Dalitz plot analyses
have already been done. However, notice that the threshold of the decay
$\phi \to KK$ is about 25 MeV away from the $\phi$ mass, so that the
available phase space is not as narrow as in the $D^\ast$ case.
Performing a Monte Carlo simulation for the decay $D\to \phi\pi$ similar
to that described above for the process $B\to \bar D^\ast D_s$, we have
found that the expected effect could be in this case as large as 30\%.

\acknowledgements

We are grateful to D.\ Cinabro, C.\ G\"obel and G.\
Gonz\'alez--Sprinberg for interesting discussions and comments. D.G.D.\
acknowledges financial aid from Fundaci\'on Antorchas (Argentina). This
work has been partially supported by CONICET and ANPCyT (Argentina).

\begin{table}
\begin{center}
\begin{tabular}{|c||c|c|c|c||c|}
~~~~~$\ \ n\ \ $ ~~~~~ &  0.1 MeV  & 0.5 MeV &  1 MeV & 2 MeV & fictitious\\
    \hline \hline
1 & 61\% &  30\% & 18\% & 11\% & 70\% \\
3 & 69\% &  36\% & 24\% &  17\% & 90\% \\
10 & 73\% &  41\% & 30\% & 25\% & 97\% \\
30 & 75\% &  46\% & 37\% & 36\% & 99.5\% \\
\hline
\end{tabular}
\caption[]{The ratio $r$ as explained in the text, for different values of
$n$ (1 to 30) and $\Gamma$ (0.1 to 2~MeV). The last column shows the
values of $r$ for a fictitious resonance with mass 2.6 GeV ---see text.}
\label{tab1}
\end{center}
\end{table}

\begin{figure}
\centerline{\psfig{figure=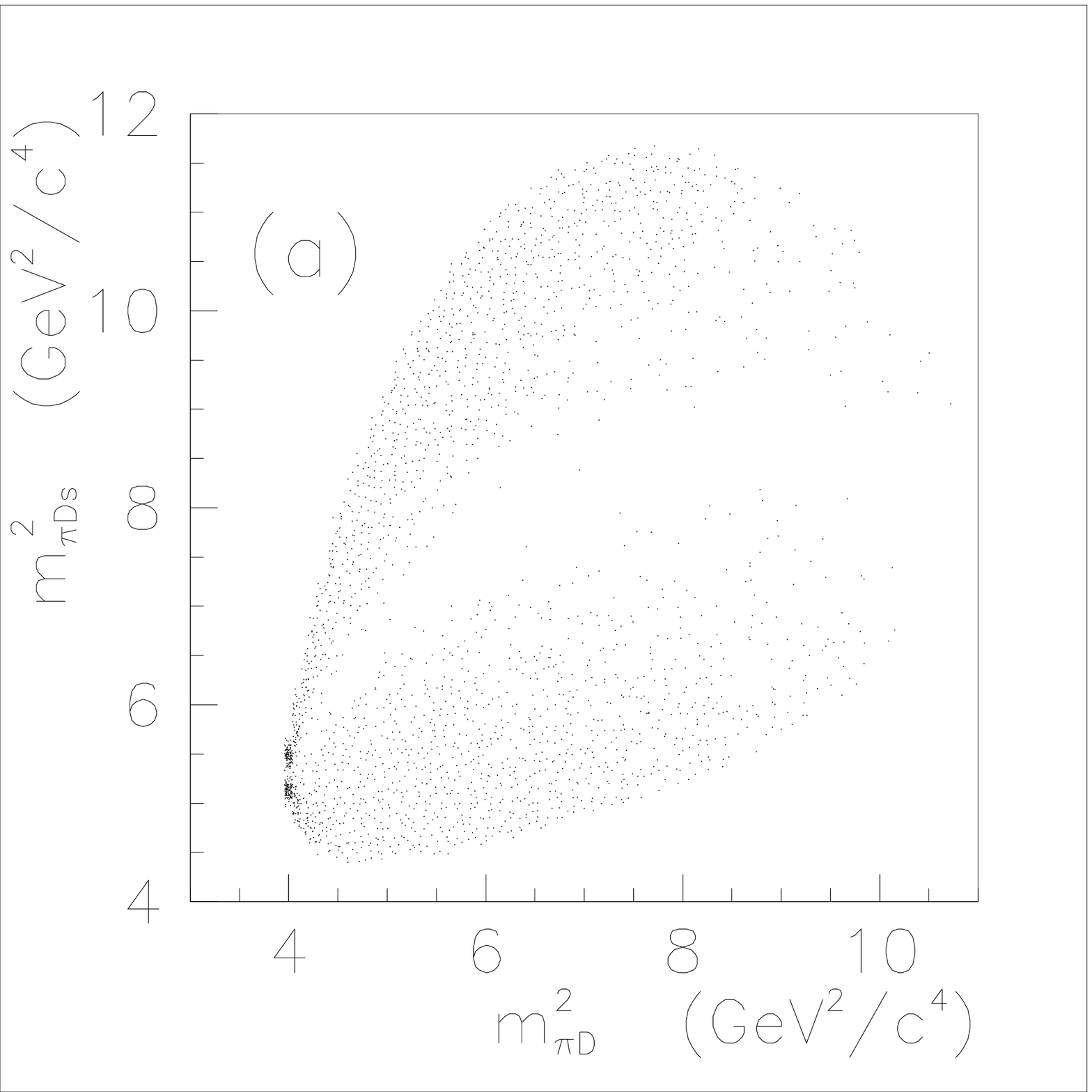,height=5cm}
\hspace{.5cm} \psfig{figure=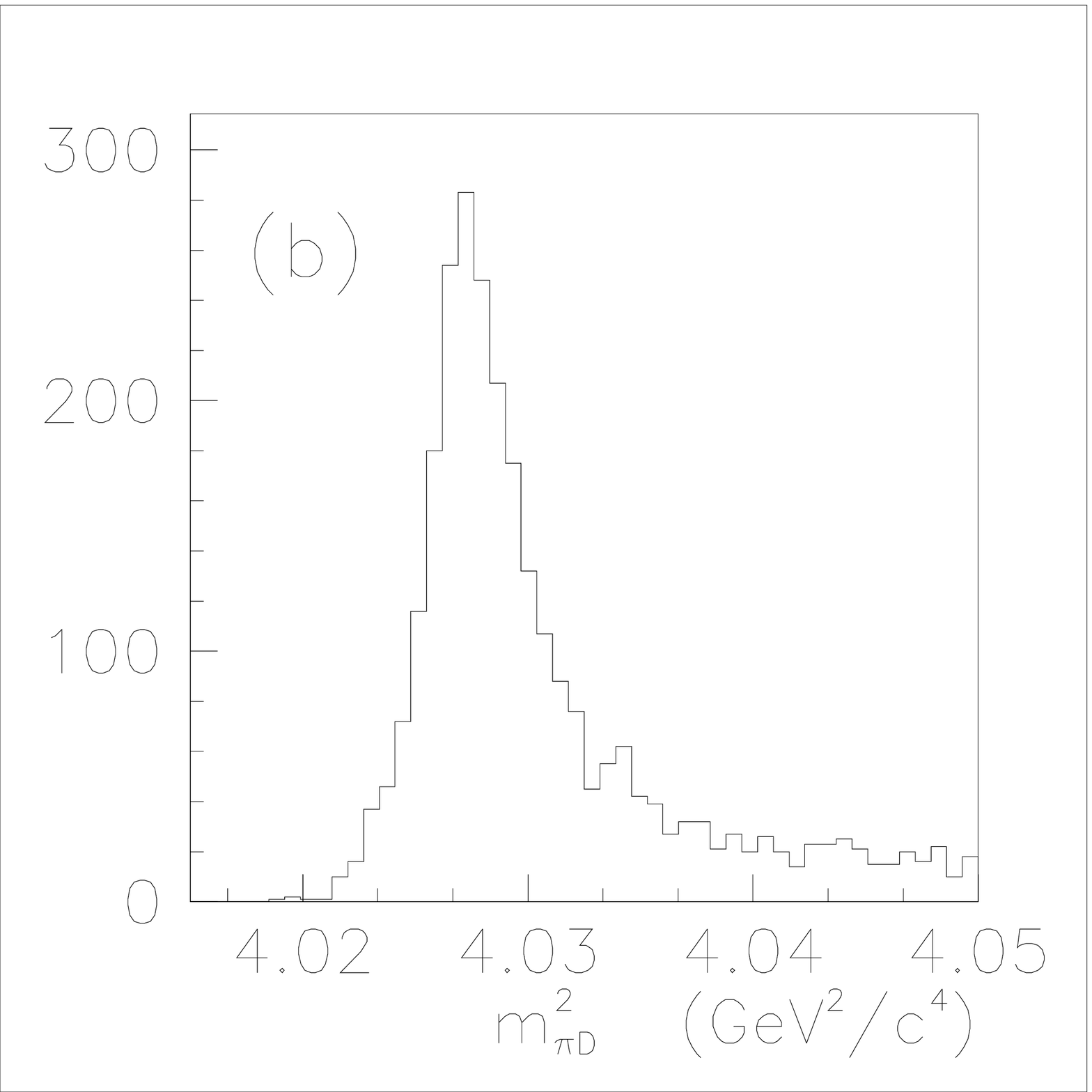,height=5cm}
\hspace{.5cm} \psfig{figure=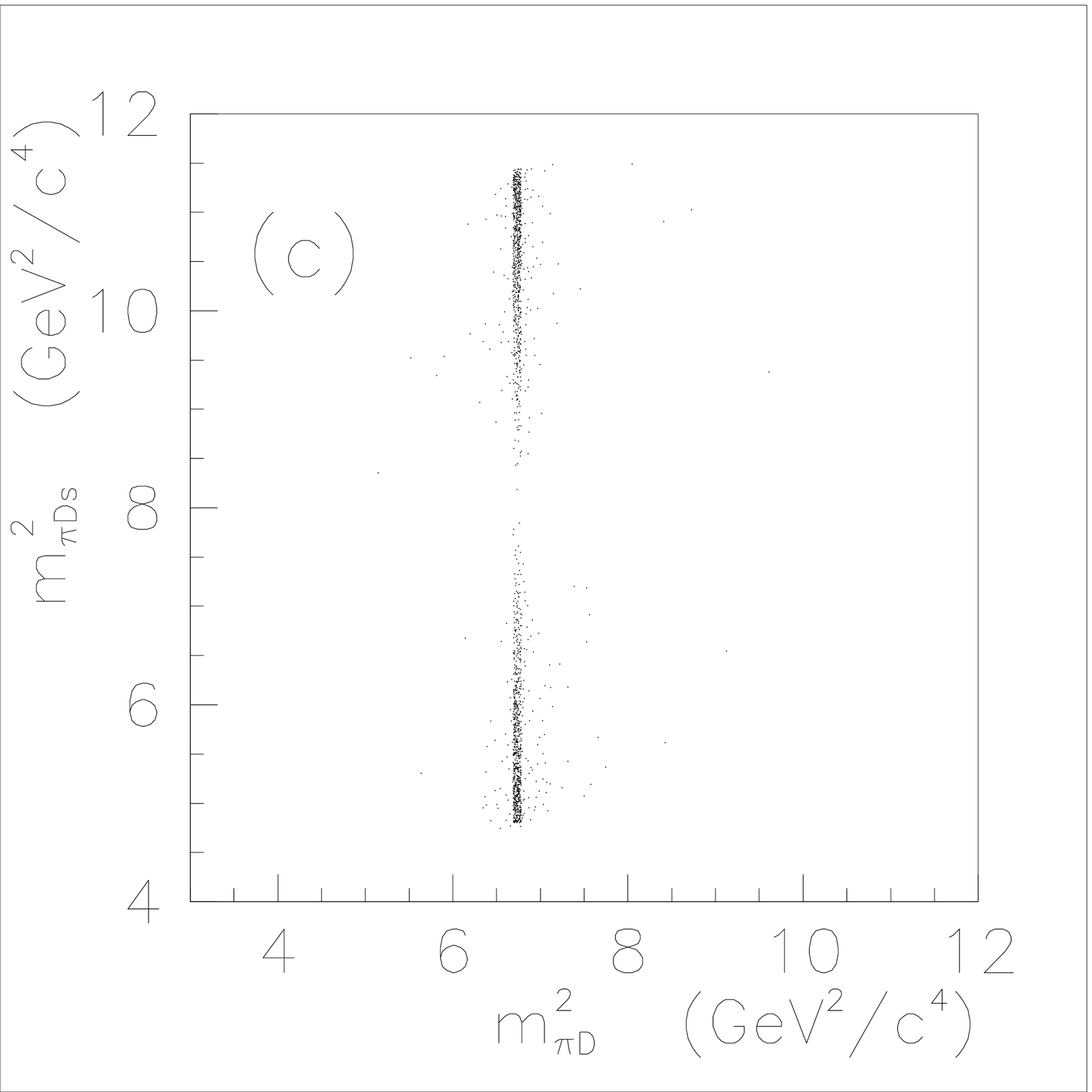,height=5cm}}
\vspace{0.3cm}
\protect\caption{
Simulation of the decay $B^+ \to \bar D^{\ast 0} D^+_s;$ $\bar
D^{\ast 0} \to \bar D^0\pi^0$, with $\Gamma = 1$~MeV and 10000 generated
events. (a) is the Dalitz plot in the plane $[(p_{\bar
D^0}+p_{\pi^0})^2$, $(p_{D^+_s}+p_{\pi^0})^2]\,$, whereas (b) is a
projection on the $(p_{\bar D^0}+p_{\pi^0})^2$ axis. (c) is same as (a),
with the $\bar D^{\ast 0}$ mass shifted to 2.6~GeV}
\end{figure}

\end{document}